\documentstyle[prl,preprint,aps]{revtex}
\input epsf

\def\vct#1{{\bf #1}}
\def\vct#1{{\bm #1}}
\newfam\bmfam
\newfont{\tenbm}{cmmib10 scaled1000}
\newfont{\eigbm}{cmmib10 scaled800}
\newfont{\sevbm}{cmmib10 scaled700}
\newfont{\sixbm}{cmmib10 scaled600}
\newfont{\fivbm}{cmmib10 scaled500}
\textfont\bmfam\tenbm
\scriptfont\bmfam\sevbm
\scriptscriptfont\bmfam\fivbm
\def\bm{\fam\bmfam\tenbm}
\let\tenfootnote=\footnote
\def\footnote{%
  \textfont\bmfam\eigbm
  \scriptfont\bmfam\sixbm
  \tenfootnote}

\tightenlines
\begin{document}
\preprint{hep-th/9712237 \hspace{9.5cm} CHIBA-EP-102 }
\title{Back Reaction to Rotating Detector}

\author{Suga Takayuki
 \footnote{e-mail address:psuga@cuphd.nd.chiba-u.ac.jp}}
 \address{Graduate School of Science and Technology, Chiba University,\\
          1-33 Yayoicho, Inage-ku, Chiba 263-8522,  Japan ,}
\author{Mochizuki Riuji
 \footnote{e-mail address:rjmochi@tdc.ac.jp}\\
        and \\
        Ikegami Kenji
 \footnote{e-mail address:kikegami@tdc.ac.jp}}
 \address{Laboratory of Physics, Tokyo Dental College,\\
          1-2-2 Masago, Mihama-ku, Chiba 261-8502, Japan}

\date{\today}
\maketitle
\begin{abstract}

It has been a puzzle that rotating detectors may respond
even in the appropriate vacuum defined via canonical quantization.
We solve this puzzle by taking back reaction of the detector into account.
The influence of the back reaction,
even in the detector's mass infinite limit,
appears in the response function.
It makes the detector possible to respond in the vacuum
if the detector is rotating,
though the detector in linear uniform motion never respond
in the vacuum as expected from Poincar\'e invariance.

\end{abstract}


\vfill\eject

\narrowtext

It is known that there are only two kinds
of vacua appropriate for
stationary (not necessarily static) coordinate systems
defined on the flat spacetime\cite{LP};
they are Minkowski vacuum and Fulling vacuum,
which exist on Hilbert spaces being not unitarily equivalent to each other.
The vacuum for the coordinate system uniformly rotating
relatively to the inertial frame (Minkowski frame),
to say the rotating vacuum,
is just equivalent to Minkowski (inertial) vacuum
on the region where the rotating coordinate system is defined.
The observer in
the coordinate system adapted to Fulling vacuum
regards Minkowski vacuum as a (thermal) heat bath,
which is Fulling-Davies-Unruh (FDU) thermal bath.
To examine this phenomenon,
called Unruh effect\cite{Fulling}\cite{BD}\cite{Takagi},
it is necessary that the system including a detector
model is studied \cite{DeWitt}\cite{Unruh}\cite{UW}.
The detector is named Unruh-DeWitt detector.
The response function of the detector set rest in a frame is evaluated to
see which vacuum is appropriate for the frame.
Besides, the rotation has been investigated in some papers,
where exists considerable confusion about the energetics
of detector response
\cite{UW}\cite{The_rotating}\cite{A_rotating}\cite{takagi2}\cite{GO}.
For uniform circular motion Letaw and Pfautsch\cite{LP} have pointed
out the possibility that the detector may respond
even in the state which a rotating observer regards as the vacuum.
In this paper
we shall resolve the above confusion,
that is,
we show that the rotating detector may respond
even in the vacuum in accordance with the energetics.

Firstly,
we quickly review conventional calculation of the response function.
For simplicity,
we treat a massless scalar field $\phi$.
The expansions of $\phi$ in two coordinate systems $S$ and $S'$ are
\begin{eqnarray}
 \phi (x)&=&\hspace{8pt}\sum_{\vct k}
            \Bigl\{ a_{\vct k}u_{\vct k}(x)
                   +a^\dagger_{\vct k}u_{\vct k}^*(x)\Bigr\}\; ,
 \label{ModeExGen} \\
 \phi (x')&=&\sum_{\vct k'}
             \Bigl\{ b_{\vct k'}{\cal U}_{\vct k'}(x')
                    +b^\dagger_{\vct k'}{\cal U}_{\vct k'}^*(x')\Bigr\}\; ,
 \nonumber
\end{eqnarray}
where $u$ (${\cal U}$) is the mode function
in the systems $S$ ($S'$) and
$a^\dagger_{\vct k}$ ($b^\dagger_{\vct k'}$) and $a_{\vct k}$ ($b_{\vct k'}$)
are creation and annihilation operators respectively.
$x$ ($x'$) denotes the coordinates of the system $S$ ($S'$) and
$\vct k$ ($\vct k'$)
the 3-(generalized)momentum of an $a$($b$)-particle.
The vacua $|0\rangle$ in $S$ and $|0'\rangle$ in $S'$
are defined as
\[
 a_{\vct k}|0\rangle =0 \;\;\;
 \mbox{and} \;\;\;
 b_{\vct k'}|0'\rangle =0 \; .
\]
The relational expressions among the operators
known as Bogoliubov transformation, are defined as
\begin{equation}
 \begin{array}{cccc}
  \;\;\;\;\;\;b_{\vct k'}&=&
  \displaystyle{\sum_{\vct k}\Bigl\{\alpha _{\vct k'\vct k}a_{\vct k}
                +\beta _{\vct k'\vct k}a_{\vct k}^\dagger\Bigr\}}&, \\
 \noalign{\vskip7pt}
  \;\;\;\;\;\;b_{\vct k'}^\dagger &=&
  \displaystyle{\sum_{\vct k}\Bigl\{\beta _{\vct k'\vct k}^*a_{\vct k}
                +\alpha _{\vct k'\vct k}^*a_{\vct k}^\dagger\Bigr\}}&.
 \end{array}
 \label{opBog}
\end{equation}
These coefficients $\alpha$ and $\beta$ are called Bogoliubov coefficients.
If and only if $\beta\neq 0$,
the vacuum on the system $S$ is not hte vacuum on the system $S'$:
$|0\rangle\neq|0'\rangle$.

In this paper, we consider the situation that the Unruh-DeWitt detector
on the ground state excites into the upper state
with emitting an $a$-particle in the vacuum $|0\rangle$.
Then the response function ${\cal F}(\Delta E)$ is
\begin{equation}
 {\cal F}(\Delta E)
  =\int d\tau_1\!\int d\tau_2\;e^{-i\Delta E(\tau_2-\tau_1)}G^+(x_2,x_1) \; ,
 \label{Forg}
\end{equation}
where $\Delta E$ is the energy gap between the ground and upper states and
$G^+$ is the positive frequency Wightman Green function.
$\tau_1$ and $\tau_2$ are points on the detector's proper time $\tau$,
and
\[
 x_1=x(\tau_1) \; , \; x_2=x(\tau_2) \; .
\]
Using the following relations,
\[
 \begin{array}{cccc}
 u_{\vct k}&=&\displaystyle{\sum_{\vct k'}\Bigl\{
              \alpha _{\vct k\vct k'}^*{\cal U}_{\vct k'}
              +\beta _{\vct k\vct k'}{\cal U}_{\vct k'}^*\Bigr\}}&, \\
 \noalign{\vskip7pt}
 u_{\vct k}^*&=&\displaystyle{\sum_{\vct k'}\Bigl\{
              \beta _{\vct k\vct k'}^*{\cal U}_{\vct k'}
              +\alpha _{\vct k\vct k'}{\cal U}_{\vct k'}^*\Bigr\}}&,
 \end{array}
\]
$G^+$ is expressed as
\begin{eqnarray}
 G^+(x_2,x_1)&\equiv&\langle 0|\phi (x_2)\phi (x_1)|0\rangle
 \nonumber \\
         &=&\sum_{\vct k}u_{\vct k}(x_2)u_{\vct k}^*(x_1)
 \label{GinS} \\
         &=&\sum_{\vct p}\sum_{\vct p'}\sum_{\vct k}
             \Bigl\{\alpha_{\vct k\vct p}^*\alpha_{\vct k\vct p'}
                    {\cal U}_{\vct p}(x'_2){\cal U}_{\vct p'}^*(x'_1)
                   +\alpha_{\vct k\vct p}^*\beta_{\vct k\vct p'}^*
                    {\cal U}_{\vct p'}(x'_2){\cal U}_{\vct p'}(x'_1) \Bigr.
 \nonumber \\
          && \hspace{60pt} \Bigl.
                   +\beta_{\vct k\vct p}\alpha_{\vct k\vct p'}
                    {\cal U}_{\vct p}^*(x'_2){\cal U}_{\vct p'}^*(x'_1)
                   +\beta_{\vct k\vct p}\beta_{\vct k\vct p'}^*
                    {\cal U}_{\vct p}^*(x'_2){\cal U}_{\vct p'}(x'_1)\Bigr\}\;,
 \label{GinS'}
\end{eqnarray}
where $x_1$ ($x_2$) and $x'_1$ ($x'_2$) are the same points
on the fixed trajectory $x(\tau)$ of the detector
in $S$ and $S'$ respectively\cite{BD}.

In the rest of this paper,
let the system $S$ Minkowski coordinate system
on the whole flat space-time (whole Minkowski manifold).
We choose Cartesian coordinates in the system $S$,
which choice is not essential in inertial frames.
Then the expression of the mode function
\begin{equation}
 u_{\vct k}(x)=\frac1{\sqrt{2(2\pi)^3\omega}}e^{-i\omega t+i\vct k\cdot\vct x}
              \sim e^{-i\omega t+i\vct k\cdot\vct x}\;\;\;\;\;
 \mbox{with}\;\;\omega\equiv\sqrt{\vct k^2}\;\;\;,
 \label{modeS}
\end{equation}
using eq(\ref{GinS}) (not eq.(\ref{GinS'})), leads
\begin{eqnarray}
 G^+(x_2,x_1)&=&\int\!\frac{d\vct k}{2(2\pi)^3\omega}\;
              e^{-i\omega(t_2-t_1)+i\vct k\cdot(\vct x_2-\vct x_1)}
 \label{WightmanBeforeInt} \\
             &=&\frac{-1}{(2\pi)^2}\frac1{(x_2-x_1)^2}
 \nonumber \\
             &\sim&\frac1{(t_2-t_1-i\epsilon)^2-|\vct x_2-\vct x_1|^2} \;,
 \label{Wightman}
\end{eqnarray}
where
\[
 t_1=t(\tau_1) \;, \; \vct x_1=\vct x(\tau_1) \;, \;
 t_2=t(\tau_2) \;, \; \vct x_2=\vct x(\tau_2) \;.
\]
The Green function (\ref{Wightman}) depends only on the difference
between $x_1$ and $x_2$.
Moreover we restrict ourselves to discussing the situation in which
the argument of $G^+$ is only $\Delta\tau\equiv\tau_2-\tau_1$
\footnote{
More generally
$G^+(\tau_2,\tau_1)=G^+(\Delta\tau ,T)=G^+_{\Delta\tau}(\Delta\tau)+G^+_T(T)$
is also allowed.
}.
In such a situation
linear uniform motion, circular uniform motion
and linear uniformly accelerated motion are, for example, available.
From eq.(\ref{Forg})
we define the response function per unit detector's proper time as
\begin{equation}
 \overline{\cal F}(\Delta E)\equiv\frac{{\cal F}(\Delta E)}{\overline{T}}
  =\int_{-\infty}^\infty d(\Delta\tau)\;
   e^{-i\Delta E\Delta\tau}G^+(\Delta\tau) \; ,
 \label{Forg/T}
\end{equation}
where
\[
 \overline{T}\equiv\int_{-\infty}^\infty dT \;\; ,
 \;\; T\equiv\frac{\tau_1+\tau_2}2 \; .
\]

Furthermore we impose another condition:
$S'$ is the coordinate system
in which the detector is rest (namely $t'\propto\tau$),
and normal particle interpretation is admitted.
On such a condition as
\[
 \left\{
  \begin{array}{cccccl}
   t_1'&=&t'(\tau_1)&=&c\tau_1& \\
  \noalign{\vskip3pt}
   \vct x_1'&=&\vct x'(\tau_1)&=&\vct x_0'&=\mbox{ constant} \\
  \noalign{\vskip3pt}
   t_2'&=&t'(\tau_2)&=&c\tau_2& \\
  \noalign{\vskip3pt}
   \vct x_2'&=&\vct x'(\tau_2)&=&\vct x_0'&=\mbox{ constant}
 \end{array}
 \right.
\]
\vspace{-20pt}
\[
 c=\mbox{constant}>0 \; ,
\]
the mode function is written as
\begin{equation}
 {\cal U}_{\vct k'}(x')={\cal U}_{\vct k'}(t',\vct x')
                =f_{\vct k'}(\vct x')\;e^{-i\omega 't'}\;.
 \label{modeS'}
\end{equation}
To designate eq.(\ref{GinS'}) we, using eq.(\ref{modeS'}),
introduce coefficient matrices $A$, $B$, $C$ and $D$:
\begin{equation}
 \begin{array}{ccll}
  \displaystyle{\sum_{\vct p}\sum_{\vct p'}A_{\vct p\vct p'}
               {\cal U}_{\vct p}\Bigl(x'(\tau_2)\Bigr)
               {\cal U}_{\vct p'}^*\Bigl(x'(\tau_1)\Bigr)}&=&
  \displaystyle{\sum_{\vct p}\sum_{\vct p'}A_{\vct p\vct p'}
   f_{\vct p}(\vct x_0')f_{\vct p'}^*(\vct x_0')\;
  e^{-i\omega c\tau_2+i\omega 'c\tau_1}}&, \\
 \noalign{\vskip7pt}
  \displaystyle{\sum_{\vct p}\sum_{\vct p'}B_{\vct p\vct p'}
               {\cal U}_{\vct p}\Bigl(x'(\tau_2)\Bigr)
               {\cal U}_{\vct p'}\Bigl(x'(\tau_1)\Bigr)}&=&
  \displaystyle{\sum_{\vct p}\sum_{\vct p'}B_{\vct p\vct p'}
   f_{\vct p}(\vct x_0')f_{\vct p'}(\vct x_0')\;
  e^{-i\omega c\tau_2-i\omega 'c\tau_1}}&, \\
 \noalign{\vskip7pt}
  \displaystyle{\sum_{\vct p}\sum_{\vct p'}C_{\vct p\vct p'}
               {\cal U}_{\vct p}^*\Bigl(x'(\tau_2)\Bigr)
               {\cal U}_{\vct p'}^*\Bigl(x'(\tau_1)\Bigr)}&=&
  \displaystyle{\sum_{\vct p}\sum_{\vct p'}C_{\vct p\vct p'}
   f_{\vct p}^*(\vct x_0')f_{\vct p'}^*(\vct x_0')\;
  e^{i\omega c\tau_2+i\omega 'c\tau_1}}&, \\
 \noalign{\vskip7pt}
  \displaystyle{\sum_{\vct p}\sum_{\vct p'}D_{\vct p\vct p'}
               {\cal U}_{\vct p}^*\Bigl(x'(\tau_2)\Bigr)
               {\cal U}_{\vct p'}\Bigl(x'(\tau_1)\Bigr)}&=&
  \displaystyle{\sum_{\vct p}\sum_{\vct p'}D_{\vct p\vct p'}
   f_{\vct p}^*(\vct x_0')f_{\vct p'}(\vct x_0')\;
  e^{i\omega c\tau_2-i\omega 'c\tau_1}}&,
 \end{array}
 \label{UU}
\end{equation}
where we assume
\begin{equation}
 \omega\equiv\omega (\vct p)=\sqrt{\vct p^2} \;\;,\;\;
 \omega '\equiv\omega (\vct p')=\sqrt{(\vct p')^2}.
 \label{w>0}
\end{equation}
If each of eqs.(\ref{UU}) depends on either $\Delta\tau$ or $T$,
$\omega =\omega '$ because $\omega ,\omega '\geq0$.
We define $F$ as sum of eqs.(\ref{UU}):
\[
 F(\Delta\tau,T)=\sum_{\vct p}\left\{
             A_{\vct p}f_{\vct p}f_{\vct p}^*e^{-i\omega c\Delta\tau}
            +B_{\vct p}f_{\vct p}f_{\vct p}e^{-i\omega cT}
            +C_{\vct p}f_{\vct p}^*f_{\vct p}^*e^{i\omega cT}
            +D_{\vct p}f_{\vct p}^*f_{\vct p}e^{i\omega c\Delta\tau}\right\}\;.
\]
When $F$ is independent of $T$
\[
 B_{\vct p}=0 \; , \; C_{\vct p}=0 \; ;
\]
\begin{equation}
 F(\Delta\tau)=\sum_{\vct p}\left\{
             A_{\vct p}\widetilde f_{\vct p}\widetilde f_{\vct p}^*
                e^{-i\omega c\Delta\tau}
            +D_{\vct p}\widetilde f_{\vct p}^*\widetilde f_{\vct p}
                e^{i\omega c\Delta\tau}\right\} \; ,
 \label{sumUU}
\end{equation}
where
\[
\begin{array}{c}
 A_{\vct p}\delta_{\vct p\vct p'}
=U_{\vct p\vct k}A_{\vct k\vct k'}U_{\vct k'\vct p'}^{-1}\;,\mbox{ etc.,}
\\ \noalign{\vskip4pt}
 \mbox{and }\;\;\widetilde f_{\vct p}=U_{\vct p\vct k}f_{\vct k}\;.
\\ \noalign{\vskip4pt}
 U_{\vct p\vct k}\mbox{: a unitary matrix .}
\end{array}
\]
Having assumed to have the same form as eq.(\ref{sumUU}),
$G^+$ (\ref{GinS'}) should be written as
\begin{equation}
 G^+(\Delta\tau)=\sum_{\vct p}\sum_{\vct p'}\left\{
                    \sum_{\vct k}\alpha_{\vct k\vct p}^*\alpha_{\vct k\vct p'}
                   {\cal U}_{\vct p}(x'_2){\cal U}_{\vct p'}^*(x'_1)
                   +\sum_{\vct k}\beta_{\vct k\vct p}\beta_{\vct k\vct p'}^*
                   {\cal U}_{\vct p}^*(x'_2){\cal U}_{\vct p'}(x'_1)\right\}\;.
 \label{G_UU}
\end{equation}
Though $\alpha_{\vct p\vct p'}=\alpha_{\vct p}\delta_{\pm\vct p\vct p'}$
and $\beta_{\vct p\vct p'}=\beta_{\vct p}\delta_{\pm\vct p\vct p'}$
are not always correct\cite{BD},
we symbolically write
\begin{equation}
 \sum_{\vct k}\alpha_{\vct k\vct p}^*\alpha_{\vct k\vct p'}
  =|\alpha_{\vct p}|^2\delta_{\vct p\vct p'} \;\; , \;\;
 \sum_{\vct k}\beta_{\vct k\vct p}^*\beta_{\vct k\vct p'}
  =|\beta_{\vct p}|^2\delta_{\vct p\vct p'} \;.
 \label{diag}
\end{equation}
Then
\begin{equation}
 G^+(\Delta\tau)=\sum_{\vct p'}\left\{
           |\alpha_{\vct p'}|^2\widetilde f_{\vct p'}\widetilde f_{\vct p'}^*
           \;e^{-i\omega 'c\Delta\tau}
          +|\beta_{\vct p'}|^2\widetilde f_{\vct p'}^*\widetilde f_{\vct p'}
           \;e^{i\omega 'c\Delta\tau} \right\} \; .
 \label{GpropAB}
\end{equation}
Hence eq.(\ref{Forg/T}) becomes
\begin{equation}
 \overline{\cal F}(\Delta E)=\sum_{\vct p'}\left\{
     |\alpha_{\vct p'}|^2\;\delta(c\omega '+\Delta E)
     +|\beta_{\vct p'}|^2\;\delta(c\omega '-\Delta E)
                                   \right\}|\widetilde f_{\vct p'}|^2 \; .
 \label{FpropAB}
\end{equation}
Either the first or second term in the right hand side survives
for a fixed $\vct p'$.
In this case since $\omega '\geq 0$ and $\Delta E>0$,
the first term vanishes.
It indicates energy conservation,
and then eq.(\ref{FpropAB}) reduces to
\begin{equation}
 \overline{\cal F}=\sum_{\vct p'}|\beta_{\vct p'}|^2
  \;|\widetilde f_{\vct p'}|^2\;\delta(c\omega '-\Delta E)\;.
 \label{FpropB}
\end{equation}
We compute the expectation value of $b$-particle number in $a$-vacuum;
\begin{eqnarray*}
 \langle 0|\sum_{\vct k'}b_{\vct k'}^\dagger b_{\vct k'}|0\rangle
 &=&\sum_{\vct k'}\sum_{\vct k}\beta_{\vct k'\vct k}^*\beta_{\vct k'\vct k}
 \;\;\;\;\;\;\;\;  \\
 &=&\sum_{\vct k'}|\beta_{\vct k'}|^2
\end{eqnarray*}
using eqs.(\ref{opBog}) and (\ref{diag}).
Because the response function should be proportional to the particle number,
eq.(\ref{FpropB}) is convincing.
When $\beta$ is zero $|0'\rangle =|0\rangle$,
so that it is natural that $\overline{\cal F}$ vanishes,
while a detector in Minkowski vacuum undergoes heat bath
when $\beta$ is not zero.

We use $\vct x=(x,y,z)$ for Cartesian coordinates or
$\vct x=(r,\theta ,z)$ for cylindrical coordinates of the inertial frame.
The expansion (\ref{ModeExGen}) of the field $\phi$ in Minkowski frame
becomes
\begin{equation}
 \phi (x(\tau ))=\int\frac{d\vct k}{\sqrt{2(2\pi )^3\omega}}
\left\{ a(\vct k)
e^{-i\omega t+i\vct k\cdot\vct x}+
a^{\dag} (\vct k)
e^{i\omega t-i\vct k\cdot\vct x}\right\}
 \label{ModeEx}
\end{equation}
\[
\omega\equiv\sqrt{\vct k^2}
\; ,\; \vct k\equiv (k_x,k_y,k_z)
\]
in the Cartesian coordinate system, or
$$
 \phi (x(\tau ))=\sum_n\int\frac{dk_rdk_z}{\sqrt{2(2\pi)^2\omega}}
\left\{ a_n(k_r,k_z)
J_n(k_rr)e^{-i\omega t+in\theta+ik_zz}+
a^{\dag} _n(k_r,k_z)
J_n(k_rr)e^{i\omega t-in\theta-ik_zz} \right\}
 \eqno{(\ref{ModeEx}')}
$$
\[
\begin{array}{l}
 \;\;\;\;\;\;\omega\equiv\sqrt{k_r^2+k_z^2} \; ,
 \\ \noalign{\vskip4pt}
 \;\;\;\;\;\;J_n\mbox{: Bessel function of the first kind}
\end{array}
\]
in the cylindrical coordinate system
. In the former case, the mode function is $u_{\vct k}$ in eq.(\ref{modeS}),
while in the latter case the mode function is
\begin{equation}
 u_{\vct k}(x)=\frac1{2\pi\sqrt{2\omega}}J_n(k_rr)e^{-i\omega t+in\theta+ik_zz}
 \;\;\;,\;\vct k\equiv (k_r,n,k_z) \; ,
 \label{modeSc}
\end{equation}
where we have adopted box normalization.

Thus we have obtained two ways in which we calculate the response function:
from eq.(\ref{GinS}) with the coordinate system $S$
and from eq.(\ref{GinS'}) with $S'$.
In both ways we perform for some examples.

For linear uniformly accelerated motion
(hyperbolic motion, called Rindler motion) with proper acceleration $\alpha$,
which motion is the typical example for $\beta\neq 0$,
the detector's trajectory
\(
 x(\tau )=(t(\tau ),x(\tau ),y(\tau ),z(\tau ))
\)
in $S$ is
\begin{equation}
 \left\{
 \begin{array}{cll}
  t(\tau )&=&\displaystyle{\frac 1\alpha\sinh\alpha\tau}\\
 \noalign{\vskip3pt}
  x(\tau )&=&\displaystyle{\frac 1\alpha\cosh\alpha\tau}\\
 \noalign{\vskip2pt}
  y,z&=&\mbox{constant}\; .
 \end{array}
 \right.
 \label{hyperbolic_trajectory}
\end{equation}
Substituting eq.(\ref{hyperbolic_trajectory})
to eqs.(\ref{Wightman}) and (\ref{Forg/T}),
the response function becomes\cite{BD}
\begin{equation}
 \overline{\cal F}=\frac{1}{2\pi}\frac{\Delta E}{e^{2\pi \Delta E\alpha}-1} \;.
 \label{ResponseHyperbolic}
\end{equation}
The appearance of the Planck factor in this response function
shows that the detector is immersed FDU thermal bath.
Computing the response function with Rindler coordinate system
($S'$, where $t'=\tau$ i.e. $c=1$),
we obtain the result in the form of eq.(\ref{FpropB}) via eq.(\ref{FpropAB})
\footnote{
See \S4.1
in \cite{Takagi}.
}.

For linear uniform motion,
the detector's trajectory
\(
 x(\tau )=(t(\tau ),\vct x(\tau ))
\)
in $S$ is
\begin{equation}
 \left\{
 \begin{array}{cll}
  t(\tau )&=&\gamma\tau\\
 \noalign{\vskip3pt}
  \vct x(\tau )&=&\vct Vt+\mbox{constant}=\vct V\gamma\tau+\mbox{constant ,}
 \end{array}
 \right.
 \label{linear_trajectory}
\end{equation}
where $\vct V$ is the detector's 3-velocity
and $\gamma $ is the Lorentz factor.
This is the simplest example for $\beta =0$.
The straightforward calculation yields
$\overline{\cal F}=0$
from eq.(\ref{Wightman}).
To observe the energy conservation
we perform $\Delta\tau$ integration before performing the momentum integration
in Wightman function $G^+$.
The Wightman function (\ref{WightmanBeforeInt}) is replaced by
\begin{eqnarray}
 G^+&\sim&\int\frac{d\vct p}{\omega}
          \exp\{ -i\omega t_2+i\vct p\!\cdot\!\vct x_2\}
          \exp\{ i\omega t_1-i\vct p\!\cdot\!\vct x_1\} \nonumber \\
    &=&\int\frac{d\vct p}{\omega}\exp\{ -i\omega (t_2-t_1)
+i\vct p\!\cdot\!(\vct x_2-\vct x_1)\} \nonumber \\
    &=&\int\frac{d\vct p}{\omega}\exp\{ -i(\omega -\vct p\!\cdot\!\!\vct V)
\gamma\Delta\tau\}\; ,
 \label{WightmanLinearInfinite}
\end{eqnarray}
and hence the response function (\ref{Forg/T}) becomes
\begin{equation}
 \overline{\cal F}\sim\int\frac{d\vct p}{\omega}\;
 \delta \biggl( (\omega -\vct p\!\cdot\!\!\vct V)\gamma +\Delta E\biggr) =0\; .
 \label{ResponseLinearInfinite}
\end{equation}
This equality is due to the observation that
the argument of the $\delta$-function in
this equation is positive definite since
\(
\omega =\sqrt{\vct p^2}=|\vct p|
>|\vct p||\vct V|\geq\vct p\!\cdot\!\!\vct V
\)
and $\Delta E>0$;
the transition is forbidden by energy conservation law.
This result is also obtained by the calculation with $S'$
in the same way as the above.
In this case,
$S'$ is another inertial frame,
so that $t'=\tau$ ($c=1$) and the mode function is
\begin{equation}
 {\cal U}_{\vct k'}(x')\sim e^{-i\omega 't'+i\vct k'\!\cdot x'}\;\;,
 \;\;\omega '\equiv\sqrt{\vct k'}\;\;.
 \label{modeS'l}
\end{equation}
The response function becomes
\begin{equation}
 \overline{\cal F}\sim\int\frac{d\vct p'}{\omega '}\;
 \delta (\omega '+\Delta E)=0\;\;,
 \label{ResponseLinearInfiniteS'}
\end{equation}
which relates to eq.(\ref{ResponseLinearInfinite})
taking account of Lorentz transformation.
Eq.(\ref{ResponseLinearInfiniteS'}) again means
the transition is forbidden by energy conservation law,
which is convincing from Poincar\'e invariance of Minkowski vacuum.
Thus in this example the first term in
eq.(\ref{FpropAB})
vanishes and
eq.(\ref{FpropB}) is realized.

One more example is uniform circular motion
with the detector's trajectory
\(
 x(\tau )=(t(\tau ),r(\tau ),\theta (\tau ),z(\tau ))
\)
in $S$;
\begin{equation}
 \left\{
 \begin{array}{cll}
  t(\tau )&=&\gamma\tau\\
 \noalign{\vskip3pt}
  \theta (\tau )&=&\Omega t+\mbox{constant}=\Omega\gamma\tau +\mbox{constant}\\
 \noalign{\vskip3pt}
  r,z&=&\mbox{constant}
 \end{array}
 \right.
 \label{circular_trajectory}
\end{equation}
where $\Omega$ is the detector's angular velocity.
It naively seems an example for $\beta =0$, which we discuss later.
If we, using eq.(\ref{Wightman}), calculate the response function,
we obtain
\[
 \overline{\cal F}(\Delta E)=\frac{\Delta E}{2\pi\gamma^2}\neq 0 \; .
\]
It may give strange impression;
the detector in the ground state excites
with emitting a particle in the vacuum.
To study the circumstance in detail
we perform the $\Delta\tau$ integration before the momentum integration
as in the calculation of the last example.
In this case we,
using the mode expansion eq.($\ref{ModeEx}'$) and $\vct p=(p_r,l,p_z)$,
obtain the Wightman function
\begin{eqnarray}
 G^+&\sim&\sum_{\vct p}\;\frac1{\omega}\;
J_l(p_rr)\exp\{ -i\omega t_2+il\theta _2+ip_zz\}
J_l(p_rr)\exp\{ i\omega t_1-il\theta _1-ip_zz\} \nonumber \\
    &=&\sum_l\left(\sum_{p_r,p_z}
\frac{J_l(p_rr)J_l(p_rr)e^{ip_zz}e^{-ip_zz}}{\omega}\right)
\exp\Bigl\{ -i\omega (t_2-t_1)+il(\theta _2-\theta _1)\Bigr\} \nonumber \\
    &\sim&\sum_l\;\exp\Bigl\{ -i(\omega -l\Omega )
\gamma\Delta\tau\Bigr\}\; ,
 \label{WightmanCircularInfinite}
\end{eqnarray}
which leads the response function
\begin{equation}
 \overline{\cal F}\sim
 \sum_l\;\delta \biggl( (\omega -l\Omega)\gamma +\Delta E\biggr) \neq 0\; .
 \label{ResponseCircularInfinite}
\end{equation}
As opposed to 
the case of the linear uniform motion $\omega -l\Omega$ may be negative,
so that the detector can respond.
(See Appendix.)
This result is also obtained by the calculation with $S'$.
In this case, $S'$ is co-rotating frame defined in (\ref{coordTr}),
so that $t'=\gamma\tau$
($c=\gamma$) and the mode function is
\begin{equation}
 {\cal U}_{\vct k'}(x')\sim
 \frac1{\sqrt{\omega '+l'\Omega}}
 J_{l'}(k_r'r')e^{-i\omega 't'+il'\theta '+ik_z'z'} \; ,
 \label{modeS'c}
\end{equation}
where
\begin{equation}
 \omega '\equiv\sqrt{(k_r')^2+(k_z')^2}-l'\Omega \; .
 \label{omegaRotating}
\end{equation}
The response function becomes
\begin{equation}
 \overline{\cal F}\sim
 \sum_{l'}\;\delta \biggl(\omega '\gamma +\Delta E\biggr) \neq 0\; ,
 \label{ResponseCircularInfiniteS'}
\end{equation}
using eq.(\ref{FpropAB}) and $\beta =0$;
the contribution comes from the first term in eq.(\ref{FpropAB}),
in which term the argument of $\delta$-function
is NOT positive definite in this case,
so that the equation is not expressed as eq.(\ref{FpropB}).
In spite that eq.(\ref{ResponseCircularInfiniteS'}) has 
the similar form to eq.(\ref{ResponseLinearInfiniteS'})
and the latter vanishes,
the former does not vanish.
It is because the assumption in eq.(\ref{w>0}) in the general discussion
is not satisfied in eq.(\ref{omegaRotating}).
Eq.(\ref{ResponseCircularInfiniteS'}) is the same that
eq.(\ref{ResponseCircularInfinite}).

It has been a puzzle
that rotating detectors have non-zero response in Minkowski vacuum,
which seems equivalent to the rotating vacuum.
The {\it puzzle} has been discussed by some
authors\cite{LP}\cite{Takagi}\cite{The_rotating}\cite{A_rotating},
and related to the depolarization of electrons
in strage rings\cite{BL}\cite{Matsas}.
Davies, Dray and Manogue\cite{The_rotating} insist $\overline{\cal F}=0$,
while other authors try to decode $\beta\neq 0$.

Our approach differ from them.
We consider the back reaction which the detector undergoes
as the origin of the fact that
the response function eq.(\ref{ResponseCircularInfinite})
may have non-zero value regardless of $\beta$.

The detector's mass is considered implicitly infinite
in conventional derivation of eqs.(\ref{ResponseHyperbolic}),
(\ref{ResponseLinearInfinite}) and (\ref{ResponseCircularInfinite}),
so that it has been assumed that the influence of the recoil of the detector
which emits (absorbs) particles could be ignored,
namely that the detector could be treated as with no back reaction\cite{BD}.
This assumption, however, is not correct.
Our purpose is solving the {\it puzzle} by using the fact that
the influence of the back reaction
remains even in the infinite mass limit of the detector.
In order to show the influence of the back reaction explicitly,
we re-evaluate the response function by letting the detector's mass
finite thereinafter: $m$ denotes the mass in the lower level
and $m'$ in the upper.

\begin{figure}[htbp]
 \centering \leavevmode
 \epsfbox[0 0 130 50]{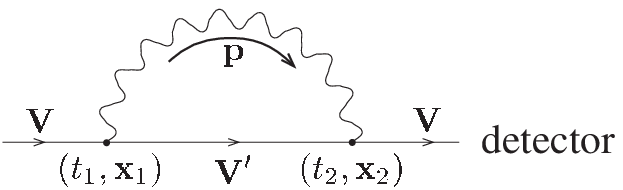}
 \caption{detector's velocity with momentum conservation (NOT Feynman diagram)}
 \label{fig}
\end{figure}

Before discussing the rotating detector with finite mass,
we recall linear uniform motion.
Consider with the coordinate system $S$ the situation in which a detector with
initial 3-velocity $\vct V$
emits a particle with
3-momentum \vct p and energy $\omega$
and the detector's 3-velocity becomes $\vct V'$
by receiving the recoil.
The equation of
the momentum conservation law is
\begin{equation}
 m\gamma \vct V=m^\prime \gamma ^\prime \vct V^\prime +\vct p\; ,
 \label{MomentumConserv1}
\end{equation}
where
\[
 \gamma \equiv \frac{1}{\sqrt{1-\vct V^2}}\; ,
 \;\;\gamma ^\prime \equiv \frac{1}{\sqrt{1-(\vct V^\prime)^2}}\; .
\]
Solving eq.(\ref{MomentumConserv1}),
\[
 \vct V^\prime = \frac{m\gamma\vct V-\vct p}
{\sqrt{(m^\prime)^2 +(m\gamma\vct V-\vct p)^2}}\; ,
\]
and then
\begin{equation}
 m^\prime \gamma ^\prime
 =\sqrt{(m^\prime)^2 +(m\gamma\vct V-\vct p)^2}\; .
 \label{MomentumConserv3}
\end{equation}
In order to obtain the Wightman function from eq.(\ref{GinS}),
we should compute the scattering amplitude
for the process shown in Fig.\ref{fig},
in which neither the detector nor the particle is virtual.
If the detector emits a particle at
$(t_1,\vct x_1)$ and absorbs it at $(t_2,\vct x_2)$,
\begin{equation}
 \vct x_2 -\vct x_1 =\vct V^\prime (t_2 -t_1)\; ,
 \label{IntermidiateVelocity}
\end{equation}
where $t_1,\vct x_1,t_2$ and $\vct x_2$ are parametrized by $\tau$,
which is the detector's proper time only in the interval $[t_1,t_2]$, as
\[
 t_1=t(\tau _1)\; ,
 \;\;\vct x_1=\vct x(\tau _1)\; ,
 \;\; t_2=t(\tau _2)\; ,
 \;\;\vct x_2=\vct x(\tau _2)\; .
\]
When the calculation is performed
in the same way as eq.(\ref{WightmanLinearInfinite})
and using eq.(\ref{IntermidiateVelocity}),
the Wightman function (\ref{GinS})
becomes
\begin{eqnarray}
 G^+ &\sim &\int\frac{d\vct p}{\omega}
\exp \{ -i\omega t_2 +i\vct p\!\cdot \!\vct x_2 \}
\exp \{ i\omega t_1 -i\vct p\!\cdot \!\vct x_1 \} \nonumber\\
  &=&\int\frac{d\vct p}{\omega}\exp \Bigl\{ -i\omega (t_2-t_1)
+i\vct p\!\cdot \!(\vct x_2 -\vct x_1)\Bigr\} \nonumber\\
  &=&\int\frac{d\vct p}{\omega}
\exp \Bigl\{ -i(\omega -\vct p\!\cdot \!\!\vct V^\prime )
\gamma ^\prime\Delta\tau\Bigr\}\;\;,
 \label{WightmanLinearFinite}
\end{eqnarray}
where
\[
 \Delta\tau = (\gamma ^\prime )^{-1}(t_2-t_1)\neq (\gamma )^{-1}(t_2-t_1)\; .
\]
$\tau$ is the time in the detector's rest frame
during the interval $[t_1,t_2]$,
and not the detector's proper time in the other intervals,
because the detector's velocity varies discretely at $t_1$ and $t_2$.
We recall that $\Delta E$ is defined as the difference
between two eigenvalues of the detector's Hamiltonian,
the generator for $\tau$-translation, before and after the emission.
This eigenvalue before the emission is not the rest mass $m$
but contains the kinetic energy,
so that $\Delta E$ is no longer the detector's energy gap:
\[
 \Delta E = m^\prime - m \widetilde{\gamma}\;,
\]
where $\widetilde{\gamma}$ is `the relative Lorentz factor'
defined as
\[
 \widetilde{\gamma}\equiv\frac 1{\sqrt{1-\widetilde{\vct V}^2}}
=\gamma\gamma^\prime (1-\vct V\!\cdot\!\vct V^\prime)\;,
\]
\vspace{-9pt}
\[
 \widetilde{\vct V}
\equiv\frac{\vct V-\vct V^\prime}{1-\vct V\!\cdot\!\vct V^\prime}\;.
\]
$\widetilde{\vct V}$ is the detector's relative velocity.
Then we redefine the energy gap $\Delta m$ as
\[
 \Delta m \equiv m^\prime - m > 0 \; .
\]
The Wightman function (\ref{WightmanLinearFinite})
gives the response function
\begin{equation}
 \overline{\cal F}\sim\int\frac{d\vct p}{\omega}\;\delta\biggl(\left(\omega 
-\vct p\!\cdot\!\!\vct V^\prime\right)\gamma '+\Delta E\biggr)\;,
 \label{ResponseLinearFinite}
\end{equation}
where, from eq.(\ref{MomentumConserv1}),
\begin{eqnarray*}
 \vct p\!\cdot\!\!\vct V^\prime
 &=& (m\gamma\vct V -m^\prime \gamma^\prime \vct V^\prime)
     \!\cdot\!\!\vct V^\prime
 \nonumber \\
 &=& m\gamma -m^\prime \gamma^\prime -m\gamma (1-\vct V\!\cdot\!\vct V^\prime )
    +\frac{m^\prime}{\gamma^\prime}\; .
\end{eqnarray*}
Then
\begin{eqnarray}
 \overline{\cal F}&\sim&   
\int\frac{d\vct p}{\omega}\;\delta \biggl(
(\omega +m^\prime \gamma ^\prime -m\gamma )\gamma ^\prime \biggr)\; .
 \label{ResponseLinearFinite2}
\end{eqnarray}
This shows the energy conservation law;
$m'\gamma '-m\gamma$ is the difference between detector's
initial mass energy and that after the emission.
The argument of the
$\delta$-function is proved positive definite in the following way.
From eq.(\ref{MomentumConserv3})
\begin{equation}
 (m'\gamma ')^2 -(m\gamma -\omega)^2
 =2m\gamma (\omega -\vct p\!\cdot\!\!\vct V)+2m\Delta m+(\Delta m)^2 \;,
 \label{C}
\end{equation}
where the second and third terms are positive definite by definition and
the first term in the right hand side is shown positive definite
in the similar way to (\ref{ResponseLinearInfinite}),
that is
\[
 (m'\gamma ')^2 -(m\gamma -\omega)^2 > 0 \; .
\]
Thus, using $m'\gamma '>0$,
\[
 m'\gamma '-m\gamma +\omega >0 \; ,
\]
which inequality shows
the response function ($\ref{ResponseLinearFinite2}$)
vanishes.

In the infinite mass limit
($m\to\infty$, $m'\to\infty$ but keeping $m'-m=\Delta m\neq 0$),
\[
 \vct V'=\vct V\;,\;\widetilde{\vct V}=0\;,
 \;\gamma '=\gamma\;,\;\widetilde{\gamma}=1\;,
 \;\Delta E=\Delta m\;.
\]
Then eq.(\ref{MomentumConserv3}) is rewritten as
\[
 m'\gamma '=m\gamma -\vct p\!\cdot\!\!\vct V+\frac{\Delta m}{\gamma '}\; ,
\]
which yields the response function in the infinite mass limit
\[
 \overline{\cal F}\sim\int\frac{d\vct p}{\omega}\;
\delta \biggl((\omega -\vct p\!\cdot\!\!\vct V)\gamma +\Delta E\biggr)\; .
\]
This coincides with eq.(\ref{ResponseLinearInfinite}).
This fact shows $\vct p\!\cdot\!\!\vct V$ in eq.(\ref{ResponseLinearInfinite})
appears due to the influence of the back reaction remaining even in
the infinite mass limit.

When the back reaction is taken into account,
$S'$ is the coordinate system co-moving before the emission and after the absorption.
We calculate with $S'$ to obtain the same response function
\[
 \overline{\cal F}\sim\int\frac{d\vct p'}{\omega '}\;
                      \delta\biggl((\omega '+m'-m\widetilde{\gamma})\biggr)
\]
as eq.(\ref{ResponseLinearFinite2}).
In the infinite mass limit
\begin{equation}
 \overline{\cal F}\sim\int\frac{d\vct p'}{\omega '}\;\delta(\omega '+\Delta m)
 \;,
 \label{FinfS'}
\end{equation}
which is equivalent to eq.(\ref{ResponseLinearInfiniteS'}).
The response function vanishes because the argument of the $\delta$-function
is clearly positive definite,
which is natural in the sense of physics.

We study uniform circular motion whose radius is $r$ (constant).
There is an essential difference between linear motion and circular motion;
the latter has no Poincar\'e invariance.
This is the key to solve the {\it rotating detector puzzle}.
We consider the situation that a detector with
initial angular velocity $\Omega$ emits a particle with
angular momentum $l$ and energy $\omega$ in circumferential direction
and the detector's angular velocity becomes
$\Omega '$ by receiving the recoil. The equation of
the angular momentum conservation law is
\begin{equation}
 m\gamma r^2\Omega =m^\prime \gamma ^\prime r^2\Omega ^\prime +l\; ,
 \label{AngularMomentumConserv1}
\end{equation}
where
$$
 \gamma \equiv \frac{1}{\sqrt{1-(r\Omega )^2}}\; ,
 \;\;\gamma ^\prime \equiv \frac{1}{\sqrt{1-(r\Omega ^\prime)^2}}\; .
$$
Solving eq.(\ref{AngularMomentumConserv1}),
\[
 \Omega ^\prime = \frac{m\gamma r^2 \Omega -l}
{r\sqrt{(m^\prime)^2 r^2 +(m\gamma r^2\Omega -l)^2}}\; ,
\]
and then
\begin{equation}
 m^\prime \gamma ^\prime
 = \frac{\sqrt{(m^\prime)^2 r^2 +(m\gamma r^2\Omega -l)^2}}r\; .
 \label{AngularMomentumConserv3}
\end{equation}
If the detector emits a particle at $(t_1,r,\theta _1,z)$
and absorbs it at $(t_2,r,\theta _2,z)$,
\begin{equation}
 \theta _2 -\theta _1 =\Omega ^\prime (t_2 -t_1)\;,
 \label{IntermidiateAngularVelocity}
\end{equation}
where $t_1,\theta _1,t_2$ and $\theta _2$ are parametrized as
\[
 t_1=t(\tau _1)\; ,
 \;\theta _1=\theta (\tau _1)\; ,
 \;t_2=t(\tau _2)\; ,
 \;\theta _2=\theta (\tau _2)\; .
\]
In the similar way to the calculation of the linear uniform motion, we obtain
\begin{equation}
 \overline{\cal F}\sim\sum_l\;
  \delta \biggl((\omega -l\Omega ')\gamma '+\Delta E\biggr)\; ,
 \label{ResponseCircularFinite}
\end{equation}
where
\[
 \Delta E = m^\prime - m \widetilde{\gamma}\; ,
\]
\[
 \widetilde{\gamma}\equiv\frac1{\sqrt{1-(r\widetilde{\Omega})^2}}
=\gamma\gamma '(1-r^2\Omega\Omega ')\; ,
\]
\[
 \widetilde{\Omega}\equiv\frac{\Omega -\Omega '}{1-r^2\Omega\Omega '}\; .
\]
Then, the response function (\ref{ResponseCircularFinite}) is rewritten as
\begin{equation}
 \overline{\cal F}\sim\sum_l\;\delta \biggl(
(\omega +m^\prime \gamma ^\prime -m\gamma)\gamma ^\prime \biggr)\; ,
 \label{ResponseCircularFinite2}
\end{equation}
which explicitly shows the energy conservation law.
In order to evaluate the argument of the $\delta$-function in this equation,
we calculate in the same way as the case of the linear uniform motion;
\[
  (m'\gamma ')^2 -(m\gamma -\omega)^2
 =2m\gamma (\omega -l\Omega)+2m\Delta m+(\Delta m)^2 \;.
\]
This equation and eq.(\ref{C}) are alike in appearance
but quite different in nature.
Though the second and third terms in the right hand side
are positive definite,
it can be zero as a whole
when $l$ in the first term increases.
Therefore $\overline{\cal F}$ may be non-zero.
In the infinite mass limit,
the response function
becomes
\begin{equation}
 \overline{\cal F}\sim\sum_l\;\delta
  \biggl((\omega -l\Omega)\gamma +\Delta E\biggr)\; .
 \label{ResponseCircularInfinite2}
\end{equation}
This is identical with eq.(\ref{ResponseCircularInfinite}).
As discussed in the case of the linear uniform motion,
$m'\gamma '-m\gamma$ in eq.(\ref{ResponseCircularFinite2}) is
the difference between the detector's initial mass energy and
that after the emission.
Hence, we consider
$l\Omega$ in eq.(\ref{ResponseCircularInfinite})
as the influence of the back reaction remaining even
in the infinite mass limit.
In other words,
{\em the detector has non-zero response because of the back reaction}.

The coordinate system $S'$ in the frame which co-rotates
before the emission and after the absorption is defined in eq.(\ref{coordTr}).
This coordinate system is, however, not well-defined out of the region
$r<1/\Omega$ as opposed to the system $S$,
which is defined in the whole space-time.
We accordingly make the boundary surface of a cylinder the radius $R$
of the base of which does not exceed $1/\Omega$.
Then the mode function is
${\cal U}_{\vct k'}(x')\sim J_{n'}(k_r'r')e^{-i\omega 't'+in'\theta '+ik_z'z'}$
which, with the above mentioned boundary, yields $\beta =0$.
The response function becomes
\begin{equation}
 \overline{\cal F}\sim \sum_{l'} \delta
  \biggl( (\omega '\gamma+m'\widetilde{\gamma}-m)\widetilde{\gamma}\biggr)\;,
 \label{ResponseCircularFiniteS'2}
\end{equation}
which vanishes in this case because the argument of the $\delta$-function
is positive definite as mentioned in Appendix.
This result is not inconsistent 
with the fact that eq.(\ref{ResponseCircularFinite2}) has non-zero value
since the boundary condition for ${\cal U}$ is different from that for $u$.
If the mode function ${\cal U}$ were defined in the whole space-time,
we would obtain the response function
with the same form as eq.(\ref{ResponseCircularFiniteS'2}).
In this case
the argument of the $\delta$-function
in the response function could be zero
as so in eq.(\ref{ResponseCircularFinite2}).
We can obtain the same result in any uniform rotating frame.
In the infinite mass limit, the response function becomes
\begin{equation}
 \overline{\cal F}\sim\sum_{l'}\delta (\omega '\gamma+\Delta m)
 \label{ResponseCircularInfiniteS'2}
\end{equation}
equivalent to eq.(\ref{ResponseCircularInfinite2}).
This equation looks like
the first term of eq.(\ref{FpropAB})
but it does not mean that eq.(\ref{ResponseCircularInfiniteS'2}) vanishes
because of the energy conservation law
in contrast to the linear motion.

We have shown that the influence of the back reaction,
even in the infinite mass limit, remains in
the argument of the $\delta$-function in the response function
in the cases of linear and circular uniform motions.
In spite of the influence,
as expected from Poincar\'e invariance, the response function
of the detector in linear uniform motion vanishes.
That is to say,
considering the circumstance in the inertial frame
in which the detector is rest
until emitting the particle,
the process that the detector emits the positive energy particle,
begins to move (gains kinetic energy) in this frame
and excites to the upper energy level,
is forbidden on energy conservation grounds.
On the other hand, 
the influence of the back reaction in uniform circular motion
induces the non-zero response function.
It is important for recognizing this to note that the inertial frame
co-moving with the detector does not exist. In other words,
there is no inertial frame in which the rotating detector {\it always}
gains kinetic energy by back reaction when emitting a particle.
If the detector's kinetic energy reduces,
the processes including
both the particle emission and the detector's excitation
may be allowed.
Therefore
`the puzzle of the rotating detector' is {\it no longer a puzzle}
if the back reaction is taken into account.
Detectors may respond even in the appropriate vacuum defined via
canonical quantization, as Letaw and Pfautsch have pointed it out\cite{LP},
which is
natural in the sense of the energy-(angular)momentum conservation.
If electrons in strage rings are used
as the detectors\cite{A_rotating}\cite{BL}\cite{Matsas},
it is necessary to take account of the back reaction.

\appendix
\section{}
We have adopted the so-called `box normalization'
as normalization of the modes in this paper.
In Cartesian coordinate system
the mode function is,
with the length of each edge of the box $L_x$ or $L_y$ or $L_z$,
\[
 u_{l,m,n}(x)=\frac1{\sqrt{2L_xL_yL_z\omega_{l,m,n}}}
                 e^{-i\omega_{l,m,n}t
                    +i\frac{l\pi}{2L_x}x
                    +i\frac{m\pi}{2L_y}y
                    +i\frac{n\pi}{2L_z}z}
\]
\[
 \mbox{with } \;\;\;
 \omega_{l,m,n}\equiv
 \sqrt{\Biggl(\frac{l\pi}{2L_x}\Biggr)^2
      +\Biggl(\frac{m\pi}{2L_y}\Biggr)^2
      +\Biggl(\frac{n\pi}{2L_z}\Biggr)^2} \;\; ,
\]
which becomes to eq.(\ref{modeS})
in the limit $L_x\to\infty$, $L_y\to\infty$, $L_z\to\infty$.
In this limit
\[
 \frac{l\pi}{2L_x}\to k_x \; , \;
 \frac{m\pi}{2L_y}\to k_y \; , \;
 \frac{n\pi}{2L_z}\to k_z \; \mbox{ and } \;
 \omega_{l,m,n}\to\omega \; .
\]
In cylindrical coordinate system in inertial frame
the mode function is,
with the radius and the height of the cylinder $R$ and $L$ respectively,
\begin{equation}
 u_{l,m,n}(x)=
   \frac1{\sqrt{2\pi R^2L\omega_{l,m,n}}J_{l+1}\left(\alpha_m^{(l)}\right)}
   J_l\left(\frac{\alpha_m^{(l)}}Rr\right)
   e^{-i\omega_{l,m,n}t+il\theta +i\frac{n\pi}{2L}z}
 \label{boundS}
\end{equation}
\[
 \mbox{with } \;\;\;
 \omega_{l,m,n}\equiv
 \sqrt{\Biggl(\frac{\alpha_m^{(l)}}R\Biggr)^2
      +\Biggl(\frac{n\pi}{2L}\Biggr)^2} \;\; ,
\]
where $\alpha_m^{(l)}$ is $m$-th positive Bessel zeroes in crescent order,
defined by $J_l(\alpha_m^{(l)})=0$.
This equation becomes to eq.(\ref{modeSc})
in the limit $R\to\infty$, $L\to\infty$.
In this limit
\[
 \frac{\alpha_m^{(l)}}R\to k_r \; , \;
 \frac{n\pi}{2L}\to k_z \; \mbox{ and } \;
 \omega_{l,m,n}\to\omega \; .
\]

The relation between the coordinate systems $S$ and $S'$
for the uniform circular motion is explicitly
\begin{equation}
 \begin{array}{lcr}
  \left\{
  \begin{array}{ccl}
   t'&=&\;\;\;t                                 \\
   x'&=&\;\;\;x\cos (\Omega t)+y\sin (\Omega t) \\
   y'&=&     -x\sin (\Omega t)+y\cos (\Omega t) \\
   z'&=&\;\;\;z                                 \\
  \end{array}
  \right.
 &
 \;\mbox{ or }\;
 &
  \left\{
  \begin{array}{ccl}
   t'      &=&t                \\
   r'      &=&r                \\
   \theta '&=&\theta -\Omega t \\
   z'      &=&z                \\
  \end{array}
  \right.\;\;.
 \end{array}
 \label{coordTr}
\end{equation}
The region of the coordinate system $S$ is the whole space-time,
but the region of the coordinate system $S'$ is limited within a cylinder
whose radius $r<1/\Omega$.
This condition is required by the relativistic theory. 
In $S'$ the mode function is
\begin{equation}
 {\cal U}_{l',m',n'}(x')=
   \frac1{\sqrt{2\pi R'^2L'(\omega '_{l',m',n'}+l'\Omega)}J_{l'+1}\left(\alpha_{m'}^{(l')}\right)}
   J_l\left(\frac{\alpha_{m'}^{(l')}}{R'}r'\right)
   e^{-i\omega '_{l',m',n'}t'+il\theta ' +i\frac{n'\pi}{2L'}z'}
 \label{boundS'}
\end{equation}
\[
 \mbox{with } \;\;\;
 \omega '_{l',m',n'}\equiv
 \sqrt{\Biggl(\frac{\alpha_{m'}^{(l')}}{R'}\Biggr)^2
      +\Biggl(\frac{n'\pi}{2L'}\Biggr)^2} \;\; .
\]
This equation is altered into eq.(\ref{modeS'c})
if taking the limit $R'\to\infty$ formally.
Due to the fact that $r'<1/\Omega$,
however, the limit $R'\to\infty$ is not proper.
Hence the mode function (\ref{modeS'c})
and accordingly Bogoliubov coefficients $\alpha$, $\beta$
are not well-defined on the whole space-time.

Before taking the limits $R\to\infty$,$L\to\infty$,
the response function which will become eq.(\ref{ResponseCircularInfinite})
in the limits is
\begin{equation}
 \overline{\cal F}\sim \sum_l\;
  \delta \biggl( (\omega_{l,m,n} -l\Omega)\gamma +\Delta E\biggr)\neq 0 \; .
 \label{ResponseCircularInfiniteB}
\end{equation}
Because of the theorem for the zeroes of Bessel function,
$\alpha_m^{(l)}/R$ may be smaller than
$l\Omega$ in $R>1/\Omega$ (cf.\cite{The_rotating}).
Hence eq.(\ref{ResponseCircularInfiniteB}) does not vanish,
so that eq.(\ref{ResponseCircularInfinite}) is justified.
Note that
$\omega$ in the $\delta$-function in eq.(\ref{ResponseCircularInfinite})
is different from that in eq.(\ref{ResponseLinearInfinite})
since
$\omega\equiv\sqrt{p_r^2+p_z^2}$ and $p\equiv\sqrt{p_x^2+p_y^2+p_z^2}$
lead $\omega\neq p$
in eq.(\ref{ResponseCircularInfinite}),
while $\omega =p\equiv\sqrt{p_x^2+p_y^2+p_z^2}$
in eq.(\ref{ResponseLinearInfinite}).
Hence eq.(\ref{ResponseCircularInfinite}) is not equivalent
to eq.(\ref{ResponseLinearInfinite})
in spite that $l\Omega =pV$ (by $\Omega =\frac Vr$ and $l=rp$).

If the systems $S$ and $S'$ are defined in the common region
which satisfies $r<1/\Omega$,
then $\overline{\cal F}=0$ due to $\alpha_m^{(l)}/R>l\Omega$.
Here we have used the fact that the Bogoliubov coefficient $\beta$
between the mode functions (\ref{boundS}) and (\ref{boundS'}) vanishes
if both of the mode functions obey a common boundary condition
\cite{The_rotating}.
Though it is possible to adopt rotating coordinate systems
other than eq.(\ref{coordTr}),
Letaw and Pfautsch pointed out
that the rotating coordinate system defined
on the whole space-time must not be stationary\cite{LP}\cite{A_rotating}.

\end{document}